\documentclass[prd,superscriptaddress,twocolumn,eqsecnum,showpacs]{revtex4}

\usepackage[dvips]{graphicx}
\usepackage{color}
\usepackage{latexsym}
\usepackage{amsmath}
\usepackage{amssymb}
\usepackage{time}

\begin{document}

\title{Casimir dependence of transverse distribution of pairs produced from a strong
constant chromo-electric background field}

\author{Fred Cooper}
\email{cooper@santafe.edu}
\affiliation{National Science Foundation,
   4201 Wilson Blvd.,
   Arlington, VA 22230}
\affiliation{Santa Fe Institute,
   Santa Fe, NM 87501}
\affiliation{Center for Nonlinear Studies,
   Los Alamos National Laboratory,
   Los Alamos, NM 87545}

\author{John F. Dawson}
\email{john.dawson@unh.edu}
\affiliation{Department of Physics,
   University of New Hampshire,
   Durham, NH 03824}

\author{Bogdan Mihaila}
\email{bmihaila@lanl.gov}
\affiliation{Materials Science and Technology Division,
   Los Alamos National Laboratory,
   Los Alamos, NM 87545}

\date{\today}

\begin{abstract}

The transverse distribution of gluon and quark-antiquark pairs produced from a strong constant chromo-electric field depends on two gauge invariant quantities,  $C_1=E^aE^a$ and $C_2=[d_{abc}E^aE^bE^c]^2$, as shown earlier in [G.~C.~Nayak and P.~van Nieuwenhuizen, Phys. Rev. D \textbf{71}, 125001 (2005)] for gluons and in [G.~C.~Nayak, Phys. Rev. D \textbf{72}, 125010 (2005)] for quarks. Here, we discuss the explicit dependence of the distribution on the second Casimir invariant, $C_2$, and show the dependence is at most a 15\% effect.
\end{abstract}

\pacs{11.15.-q, 
      12.38.-t, 
      25.75.-q, 
      25.75.Nq  
}

\maketitle

%
%

\section{Introduction}

Recently the transverse distribution of particle production from strong constant chromo-electric fields has been explicitly calculated in Ref.~\onlinecite{paper1} for soft-gluon production and in Ref.~\onlinecite{paper2} for quark (antiquark) production. This particle production method, originally discussed by Heisenberg and Euler~\cite{history1}, Schwinger~\cite{history2} and Weisskopf~\cite{history3}, has a long history as a model of the production of the quark gluon plasma following a relativistic heavy ion collision~\cite{flux,kluger1}.

The physical picture considered here is that of two relativistic heavy nuclei colliding and leaving behind a semi-classical gluon field which then non-perturbatively produces gluon and quark-antiquark pairs via the Schwinger mechanism~\cite{history2}. At high energy large hadron colliders, such as RHIC (Au-Au collisions at $\sqrt s$ = 200 GeV)~\cite{rhic} and LHC (Pb-Pb collisions at $\sqrt s$ = 5.5 TeV)~\cite{lhc}, about half the total center-of-mass energy, $E_\textrm{cm}$, goes into the production of a semi-classical gluon field~\cite{all,all1}, which can be thought to be initially in a Lorentz contracted disc.  The gluon field in SU(3) is described by two Casimir invariants, the first one, $C_1=E^a E^a$, being related to the energy density of the initial field, whereas the second one, $C_2=[d_{abc}E^aE^bE^c]^2$, is related to the SU(3) color hypercharge left behind by the leading particles.  So the question we want to study in this short note is how sensitive the transverse distribution is to this second Casimir invariant $C_2$. In a future paper we will discuss how the results for the  transverse distribution are modified by the back reaction problem for the chromo-electric field.  Some of the history of previous work on pair production in QCD  is found in the papers of~\cite{yildiz1,wkb,sav,schd,yildiz}.

\vfill

%
%

\section{Pair production rates in QCD by the Schwinger mechanism}

In Ref.~\onlinecite{paper1}, Nayak and Nieuwenhuizen obtained the following gauge invariant formula for the number of non-perturbative soft gluons produced per unit time and per unit volume and per unit transverse momentum from a given constant chromo-electric field $E^a$:
\begin{align}
\label{1}
   &
   \frac{\mathrm{d}N_{gg}}{\mathrm{d}t \,
   \mathrm{d}^3x \, \mathrm{d}^2p_T}
   \\ \notag &
   =
   \frac{1}{4\pi^3} \, \sum_{j=1}^3 \, |g\lambda_j| \,
   \ln
   \Bigl [ \,
      1
      +
      \exp
      \Bigl (
         - \frac{ \pi p_T^2}{|g\lambda_j|}
      \Bigr ) \,
   \Bigr ] \>.
\end{align}
Here $\lambda_j$ are real positive quantities defined as:
\begin{align}
   &~\lambda_1^2=\frac{C_1}{2}~\bigl ( 1 - \cos \theta \bigr )
   \>,
   \notag \\
   &~\lambda_2^2=\frac{C_1}{2}~\bigl [
      1 + \cos (\pi/3 -\theta) \bigr ]
   \>,
   \notag \\
   &~\lambda_3^2=\frac{C_1}{2}~\bigl [
   1 + \cos ( \pi/3+\theta) \bigr ]
   \>,
\end{align}
where $\theta$ is real and given by~\cite{typo}:
\begin{equation}
   \cos( 3 \theta ) = - 1 + 6 \, C_2/C_1^3
   \>.
\end{equation}
For gluons, the range in $\theta$ is $0 \le \theta \le 2 \pi/3 $.
The eigenvalues $\lambda_j$ depend only on the Casimir invariants for SU(3)
\begin{align}
   C_1=E^aE^a \>,
   \qquad
   C_2=[d_{abc}E^aE^bE^c]^2
   \>,
\end{align}
where a, b, c = 1,$\cdots$,8 are the color indices of the adjoint representation of the gauge group SU(3). Note that $0 \le C_1^3/(3C_2) \le 1$.

\begin{figure}[t!]
   \centering
   \includegraphics[width=0.9\columnwidth]{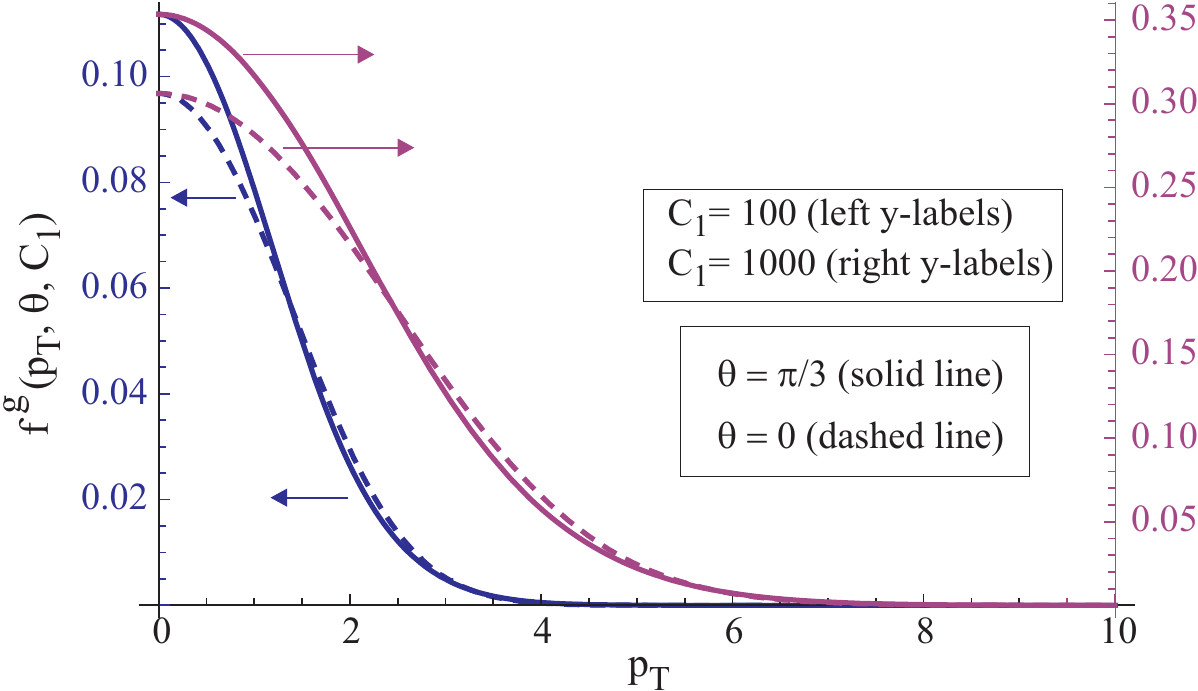}
   \caption{\label{fig1}(Color online)
   Transverse production rate for gluons for $C_1 = 100$ and 1000~GeV$^4$ and for $\theta = 0$ and $\theta = \pi/3$, as a function of~$p_\mathrm{T}$.
   For simplicity we denote here the gluon production rate given in Eq.~\eqref{1} by $f^g(p_T, \theta, C_1)$. The chosen values for $\theta$ give the minimum and maximum values of the distribution at the maximum.}
\end{figure}
\begin{figure}[b!]
   \centering
   \includegraphics[width=0.9\columnwidth]{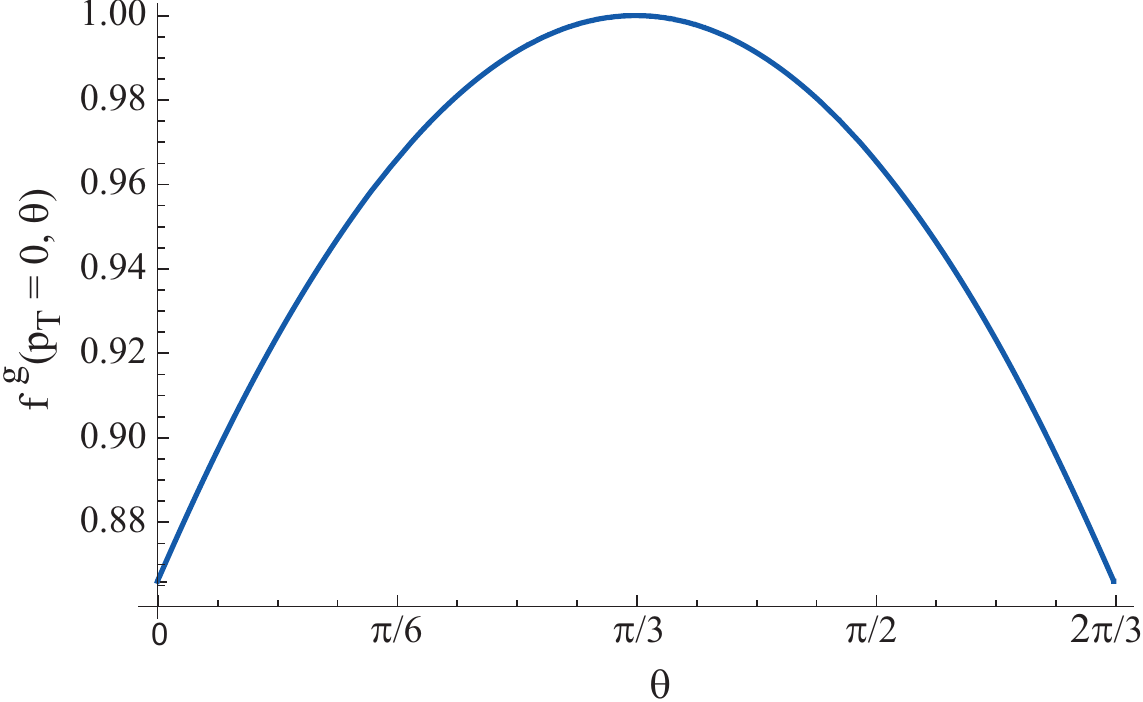}
   \caption{\label{fig2}(Color online)
   Normalized (to the maximum) forward production rate for gluons as a function of~$\theta$.}
\end{figure}

In Ref.~\onlinecite{paper2}, Nayak obtained the following gauge invariant formula for the number of non-perturbative quarks (antiquarks) pairs produced per unit time, per unit volume and per unit transverse momentum from a given constant chromo-electric field $E^a$:
\begin{align}
   &
   \frac{\mathrm{d}N_{q,\bar q}}
     {\mathrm{d}t \, \mathrm{d}^3x \, \mathrm{d}^2 p_\mathrm{T}}
   \label{2} \\
   &=
   -
   \frac{1}{4\pi^3} \, \sum_{j=1}^3 \, |g\lambda_j| \,
   \ln
   \Bigl \{ \,
      1
      -
      \exp
      \Bigl [
         - \frac{ \pi ( p_T^2+m^2 )}{|g\lambda_j|}
      \Bigr ] \,
   \Bigr \}  \>,
   \notag
\end{align}
where $m$ is the effective mass of the quark and the eigenvalues $\lambda_j$ are given by
\begin{align}
   & \lambda_1=\sqrt{\frac{C_1}{3}}~\cos \theta\,,
   \notag \\
   & \lambda_2=\sqrt{\frac{C_1}{3}}~\cos (2\pi/3-\theta)\,,
   \notag \\
   & \lambda_3=\sqrt{\frac{C_1}{3}}~\cos (2\pi/3+\theta)
   \>,
\end{align}
with $\theta$ given by
\begin{align}
   \cos^2 (3 \theta ) = 3 \, C_2/C_1^3
   \>.
\end{align}

\begin{figure}[t!]
   \centering
   \includegraphics[width=0.9\columnwidth]{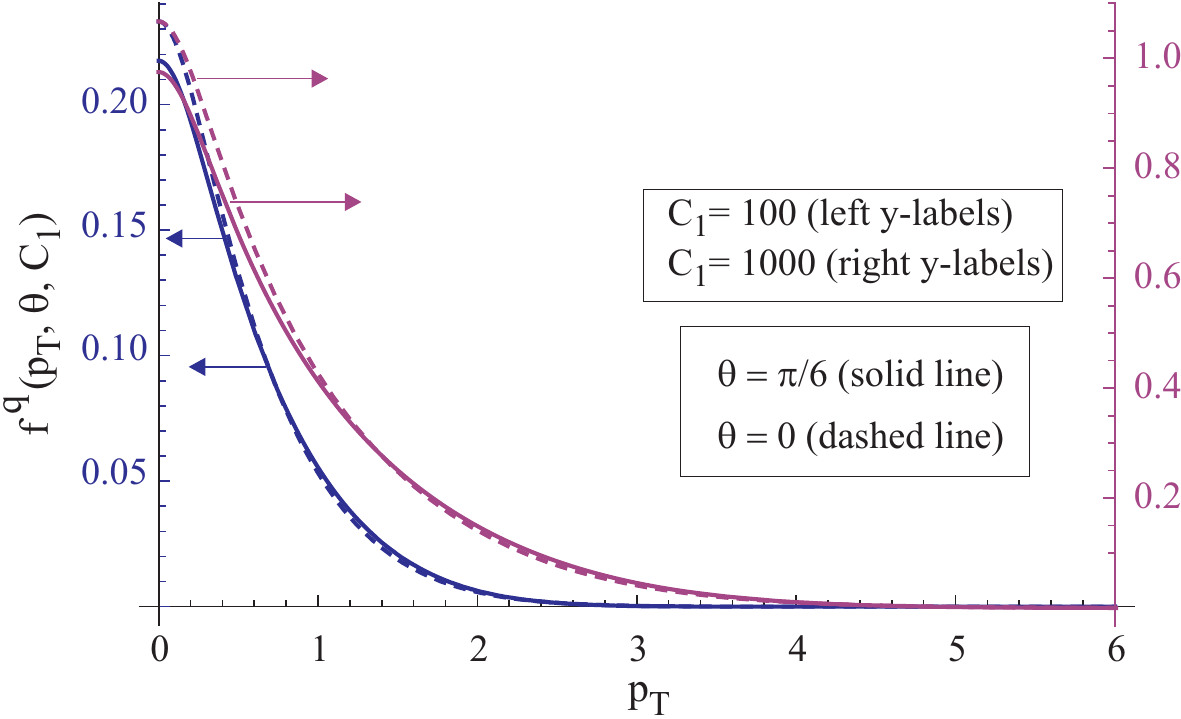}
   \caption{\label{fig3}(Color online)
   Transverse production rate for quarks for $C_1 = 100$ and 1000~GeV$^4$ for $\theta=0, \pi/6$, as a function of~$p_\mathrm{T}$.
   For simplicity we denote here the quark production rate given in Eq.~\eqref{2} by $f^q(p_T, \theta, C_1)$, The chosen values for $\theta$ give the minimum and maximum values of the distribution at the maximum. We take $m=m_q \approx 1/3$~GeV.}
\end{figure}
\begin{figure}[b!]
   \centering
   \includegraphics[width=0.9\columnwidth]{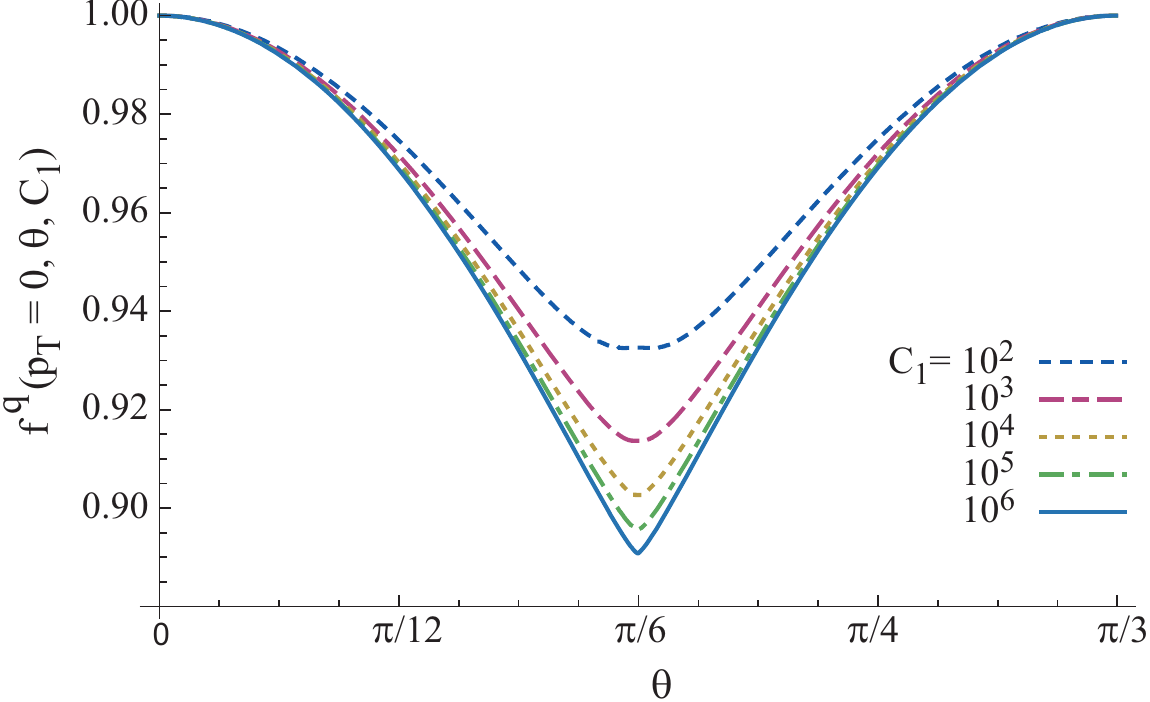}
   \caption{\label{fig4}(Color online)
   Normalized (to the maximum) forward production rate for quarks as a function of~$\theta$ and for increasing values of~$C_1$. We take $m=m_q \approx 1/3$~GeV.}
\end{figure}

For quarks, the range in $\theta$ is $0 \le \theta \le \pi/3 $, which is half the range for gluons.  The difference in sign in the logarithm between the gluon distribution and the quark distribution is related to bose \emph{vs.}\ fermi statistics, with the eigenvalues $\lambda_j$ acting  as effective temperatures.  We find that in the quark case, because of the quark mass, the forward production depends on $C_1$ and on $\theta$, whereas the normalized gluon distribution in the forward direction depends only on $\theta$ for a given initial energy density.  The value of $C_1$ can be estimated from the initial center-of-mass energy of the colliding ions, and the volume of the Lorentz contracted Nuclei.  For example for gold, $R \approx 10$~fm and at RHIC the center-of-mass energy is $\approx 200$~GeV per nucleon.  The initial density is then of the order
\begin{equation}
   \rho = E_\textrm{cm}/(V_0 \gamma) \>,
\end{equation}
with $V_0 = 4/3 \pi R^3$, and $\gamma = M_\textrm{ion}/ E_\textrm{cm}$. For the above RHIC case $\rho \approx  100$~GeV$^4$.  We take $m=m_q \approx 1/3$~GeV.

\begin{figure}[t!]
   \centering
   \includegraphics[width=0.9\columnwidth]{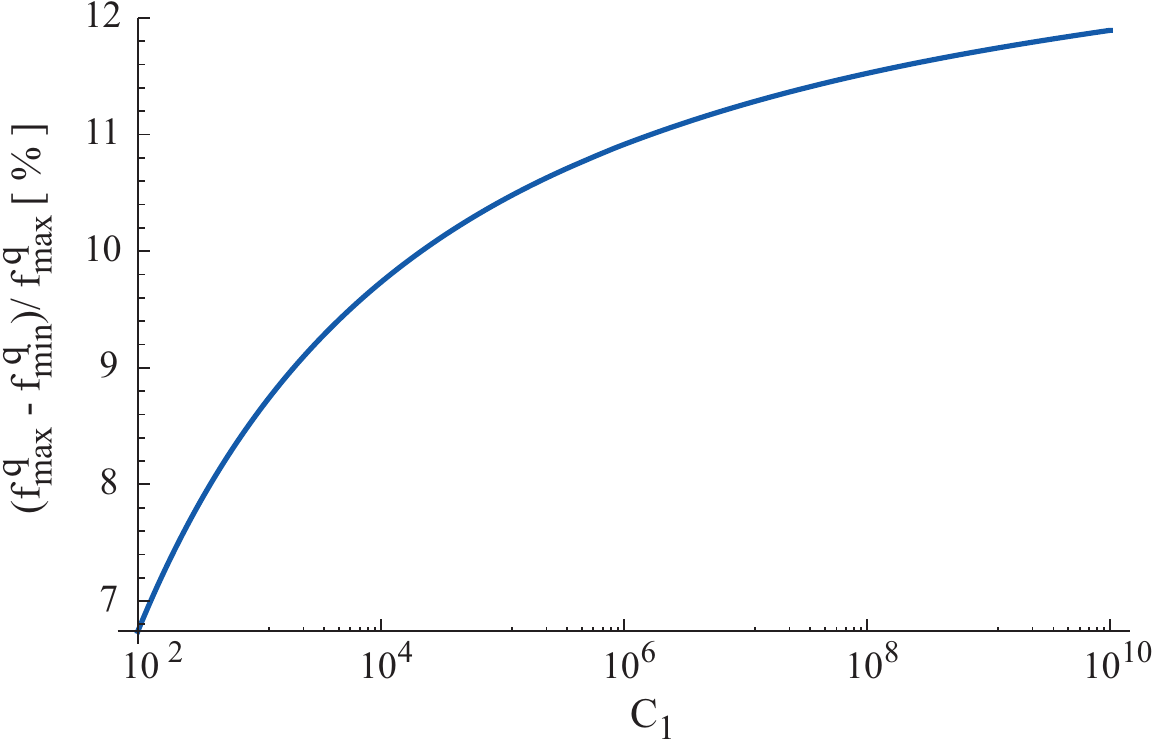}
   \caption{\label{fig5}(Color online)
   Normalized (to the maximum) percentage variation of forward production rates for quarks as a function of~$C_1$.
   For quarks, the maximum and minimum values of the forward production rate are reached for $\theta=0$ and $\theta=\pi/6$, respectively. We take $m=m_q \approx 1/3$~GeV.}
\end{figure}

%
%

\section{Results}

In Fig.~\ref{fig1} we plot the rate of gluon production as a function of the transverse momentum for $\theta$ = $0$ and $\pi/3$ for two values of the initial energy density $C_1= 100$ and 1000~GeV$^4$. These values of $\theta$ give the minimum and maximum values for the gluon production rate given in Eq.~\eqref{1}. In Fig.~\ref{fig2} we show the percentage variation of the magnitude of the normalized (to the maximum) distribution in the forward direction for gluon pair production as a function of $\theta$.  We see that this result depends only on $\theta$ because of the absence of a mass term.  The maximum variation on $\theta$ of the pair production rate occurs in the forward direction and is approximately 15\%.  We see that the maximum value of the pair production rate occurs at $\theta = \pi/3$.

In Fig.~\ref{fig3} we plot the rate of quark production as a function of the transverse momentum for $\theta$ = $0$ and $\pi/6$ for the same two values of the initial energy density, $C_1= 100$ and 1000~GeV$^4$.  These values of $\theta$ give the minimum and maximum values for the quark production rate given in Eq.~\eqref{2}.  In Fig.~\ref{fig4} we show the percentage variation of the magnitude of the normalized (to the maximum) distribution in the forward direction for quark production as a function of $\theta$.  For quarks, this quantity  depends on both $\theta$ and $C_1$ but becomes independent of $C_1$ as the initial energy density increases at which point the mass of the quark becomes irrelevant.  The maximum value of the quark production rate occurs at the endpoints $\theta=0$ and $\theta = \pi/3$.  For quarks, the maximum percentage variation, which occurs between $\theta = 0$ and $\theta= \pi/6$ is a function of~$C_1$.  This percentage variation asymptotes as a function of $C_1$ to a value of approximately 13\%, as shown in Fig.~\ref{fig5}.

%
%
\bigskip

\section{Conclusions}

We have considered the dependence of the pair production rate of quarks and gluons from a strong chromo-electric field and have discovered that the effect of the second Casimir invariant of SU(3), which was not present in the electric field problem, effects the distribution by less than 15\%.   This event by event dependence of the transverse momentum distribution of jets on $C_2$ may be something of interest at heavy ion colliders.

%
%

\begin{acknowledgments}
This work was performed in part under the auspices of the United States Department of Energy. The authors would like to thank the Santa Fe Institute for its hospitality during the completion of this work.
\end{acknowledgments}

\vfill

%
%


\begin{thebibliography}{aaaaaaa}

\bibitem{paper1} G.C. Nayak and P. van Nieuwenhuizen  Phys. Rev. D \textbf{71}, 125001 (2005).

\bibitem{paper2} G.C. Nayak, Phys. Rev. D \textbf{72}, 125010 (2005).

\bibitem{history1} W. Heisenberg and H. Euler, Z. Physik \textbf{98}, 714 (1936).
\bibitem{history2} J. Schwinger, Phys. Rev. \textbf{82}, 664 (1951).
\bibitem{history3} V. Weisskopf, Kong. Dans. Vid. Selsk. Math-fys. Medd. \textbf{XIV} No. 6 (1936);
English translation in: {\it Early Quantum Electrodynamics: A Source Book}
(Cambridge University Press, Cambridge, United Kingdom, 1994).

\bibitem{flux} A. Bialas and W. Czyzn, Phys. Rev. D \textbf{30}, 2371 (1984);
A. Bialas, W. Czyzn, A. Dyrek and W. Florkowski, Nucl. Phys. B \textbf{296}, 611 (1988);
K. Kajantie and T. Matsui, Phys. Lett. \textbf{164}B, 373 (1985);
G. Gatoff, A. K. Kerman and T. Matsui, Phys. Rev. D \textbf{36}, 114 (1987).

\bibitem{kluger1} F. Cooper and E. Mottola,  Phys. Rev. D {\bf 40}, 456 (1989), Y. Kluger, J.M. Eisenberg, B. Svetitsky, F.  Cooper and E. Mottola,
Phys. Rev. Lett. 67 (1991) 2427;
Y. Kluger, J.M. Eisenberg, B. Svetitsky, F. Cooper and E. Mottola,
 Phys. Rev. D {\bf 45} (1992)4659

\bibitem{rhic} L. McLerran and M. Gyulassy, Nucl. Phys. A \textbf{750}, 30 (2005).

\bibitem{lhc} Y. Schutz, J. Phys. G \textbf{30}, S903 (2004).

\bibitem{all} G. Baym, Phys. Lett. B \textbf{138}, 18 (1984);
K. Kajantie and T. Matsui, Phys. Lett. B \textbf{164}, 373 (1985);
K.J. Eskola and M. Gyulassy, Phys. Rev. C \textbf{47}, 2329 (1993);
G.C. Nayak and V. Ravishankar, Phys. Rev. D \textbf{55}, 6877 (1997);
Phys. Rev. C \textbf{58}, 356 (1998);
R.S. Bhalerao and G. C. Nayak, Phys. Rev. C \textbf{61}, 054907 (2000);
Y. Kluger, J.M. Eisenberg, B. Svetitsky, F. Cooper, E. Mottola, Phys. Rev. Lett. \textbf{67}, 2427 (1991);
Phys. Rev. D \textbf{45}, 4659 (1992);
D. Kharzeev and K. Tuchin, hep-ph/0501234.

\bibitem{all1} L. McLerran and R. Venugopalan Phys. Rev. D \textbf{50}, 2225 (1994);
D.D. Dietrich, G.C. Nayak and W. Greiner, Phys. Rev. D \textbf{64}, 074006 (2001);
M. Gyulassy and L. McLerran, Phys. Rev. C \textbf{56}, 2219 (1997);
D.D. Dietrich, Phys. Rev. D \textbf{70}, 105009 (2004).

\bibitem{yildiz1} M. Claudson, A. Yildiz and P.H. Cox, Phys. Rev. D \textbf{22}, 2022 (1980).

\bibitem{wkb}  A. Casher, H. Neuberger and S. Nussinov, Phys. Rev. D \textbf{20}, 179 (1979).

\bibitem{sav} S.G. Matinyan and G.K. Savvidy, Nucl. Phys. B \textbf{134}, 539 (1978).

\bibitem{schd} H-T Sato, M.G. Schmidt and C. Zahlten, hep-th/0003070;
M. Reuter, M.G. Schmidt and C. Schubert, Ann. Phys. \textbf{259}, 313 (1997).

\bibitem{yildiz} A. Yildiz and P.H. Cox, Phys. Rev. D \textbf{21}, 1095 (1980).

\bibitem{typo} We note there is a misprint in Eq.~(3) of Ref.~\cite{paper1}.
   This misprint was first noted in
   G.C.~Nayak and R.~Shrock,
   Phys. Rev. D \textbf{77}, 045008 (2008).


\end{thebibliography}
\end{document}